\documentclass[twocolumn,showpacs,preprintnumbers,amsmath,amssymb,showkeys]{revtex4}
\usepackage{graphicx}% Include figure files
\usepackage{dcolumn}% Align table columns on decimal point
\usepackage{bm}% bold math
\usepackage{epsfig}
\usepackage{amssymb}
\usepackage{natbib}

\begin{document}

\title{Cascades of Dynamical Transitions in an Adaptive Population}% Force line breaks with \\

\author{H. M. Yang, Y. S. Ting and K. Y. Michael Wong}
\email{hmyang@ust.hk, phkywong@ust.hk} \affiliation{ Department of
Physics, Hong Kong University of Science and Technology, Hong Kong,
China}
\date[]{31 August 2006}

\begin{abstract}
In an adaptive population which models financial markets and
distributed control, we consider how the dynamics depends on the
diversity of the agents' initial preferences of strategies. When the
diversity decreases, more agents tend to adapt their strategies
together. This change in the environment results in dynamical
transitions from vanishing to non-vanishing step sizes. When the
diversity decreases further, we find a cascade of dynamical
transitions for the different signal dimensions, supported by good
agreement between simulations and theory. Besides, the signal of the
largest step size at the steady state is likely to be the initial
signal.
\end{abstract}

\maketitle

\section{Introduction}

Many natural and artificial systems consist of a population of
agents with coupled dynamics. Through their mutual adaptation, they
are able to exhibit interesting collective behavior. Although the
individuals are competing to maximize their own payoffs, the system
is able to self-organize itself to globally efficient states.
Examples can be found in economic markets and communication networks
\cite{Anderson1988,Challet1997,Wei1995,Schweitzer2002}.

An important factor affecting the behavior of an adaptive population
is the dependence of the payoffs on the environment experienced by
the individual agents. The payoffs facilitate the agents to assess
the preferences of their decisions, hence inducing them to take
certain actions when they experience similar dynamical environment
in the future. Thus, the payoff function is crucial to the mechanism
of adaptation.

As a prototype of an adaptive population, the Minority Game (MG)
considers the dynamics of the buyers and sellers in a model of the
financial market, in which the minority group is the winning one
\cite{Challet1997}. A good indicator of the mutual adaptation of the
agents is the reduction of the variance of the buyer population to
values below those of random fluctuations \cite{Challet1997}.
Furthermore, this variance has a universal dependence on the
complexity of the strategies adopted by the agents, dropping to a
minimum when the complexity is reduced to a universal critical
value, and rapidly rising thereafter \cite{Savit1999,Manuca2000}.
Theoretical studies using the replica method
\cite{Challet2000,Marsili2000} and the generating functional
\cite{Heimel2001,Coolen2005} confirmed these general trends.

The agents in the original version of MG uses a \textit{step} payoff
function \cite{Challet1997,Savit1999,Manuca2000}, meaning that the
payoffs received by the winning group are the same, irrespective of
the \textit{winning margin} (the difference between the majority and
minority group). Latter versions of MG uses a \textit{linear} payoff
function \cite{Challet2000,Marsili2000,Heimel2001,Coolen2005}, in
which the payoffs increase with the winning margin. Other payoff
functions yield the same macroscopic behavior in their dependence of
the population variance on the complexity of strategies
\cite{Li2000,Lee2003}. Thus, the behavior of the population is
universal as long as the payoff function favors the minority group.
A recent extension of the MG considers payoff functions which reward
the minority agents only when they win by a large margin, but punish
them when the winning margin is small \cite{de Martino2004}. The
extended model displays a smooth crossover from a minority game to a
majority game when the payoff function is tuned.

However, when one considers details beyond the population variance,
one can find that the agents self-organize in different ways induced
by different payoff functions. For a payoff function that favors a
large winning margin, the distribution of the buyer population is
doubled-peaked \cite{Challet1997}. This shows that the dynamics of
the population self-organizes to favor large winning margins of
either the buyers or sellers, since the agents have adapted
themselves to maximize their payoffs.

In this paper, we compare the behavior of MGs using step and linear
payoffs. Previously, we found that the population variance scales as
a power law of the \textit{diversity} for a step payoff
\cite{Wong2004,Wong2005}. Diversity refers to the variance of the
initial biases of the strategy payoffs of the agents. In a
population with diverse preferences of strategies, the adaptation
rate is slow, resulting in small fluctuations of the buyer or seller
population. As we shall see, when the payoff function becomes
linear, the scaling relation between the variance and the diversity
for the step payoff is replaced by a continuous dynamical transition
from a vanishing variance at high diversity to a finite variance at
low diversity. The dynamical transition is due to the payoffs being
enhanced by large winning margins at low diversity. Furthermore, for
systems with multi-dimensional signals feeding the strategies, the
dynamical transition in each dimension do not take place at the same
transition point. Rather, there is a cascade of dynamical
transitions for the different signal dimensions. This rich behavior
demonstrates the flexibility of an adaptive population for
self-organizing to states in which agents maximize their payoffs,
and is hence important in the modeling of economics and distributed
control.

\section{The Minority Game}

The Minority Game model consists a population of \textit{N} agents
competing for limited resources, \textit{N} being odd
\cite{Challet1997}. Each agent makes a decision 1 or 0 at each time
step, and the minority group wins. For economic markets, the
decisions 1 and 0 correspond to buying and selling respectively, so
that the buyers can win by belonging to the minority group, which
pushes the price down, and vice versa. For typical control tasks
such as the distribution of shared resources, the decisions 1 and 0
may represent two alternative resources, so that less agents
utilizing a resource implies more abundance. The decisions of each
agent are responses to the environment of the game, described by
\textit{signal} $\mu^{*}(t)$ at time \textit{t}, where
$\mu^{*}(t)=1,...,D$. These responses are prescribed by
\textit{strategies}, which are binary functions mapping the
\textit{D} signals to decisions 1 or 0. In this paper, we consider
\textit{endogenous} signals, which are the \textit{history} of the
winning bits in the most recent \textit{m} steps. Thus, the
strategies have an input dimension of $D=2^{m}$, and the parameter
$\alpha\equiv\textit{D}/N$ is referred to as the
\textit{complexity}. Before the game starts, each agent randomly
picks \textit{s} strategies. Out of her \textit{s} strategies, each
agent makes decisions according to the most successful one at each
step. The success of a strategy is measured by its cumulative
payoff, as explained below.

Let $\xi_{i}^{\mu}(t)=\pm1$ when the decisions of strategy
\textit{a} are 1 or 0 respectively, responding to signal
\textit{$\mu$}. Let $a^{*}(i,t)$  be the strategy adopted by agent
\textit{i} at time \textit{t}. Then
$A(t)\equiv\sum_{i}\xi_{a^{*}(i,t)}^{\mu^{*}(t)}/N$ is the excess
demand of the game at time \textit{t}. The payoff received by
strategy $a$ is then $-\xi_{a}^{\mu^{*}(t)}\varphi(\sqrt{N}A(t))$,
where $\varphi$ is the payoff function. For step and linear payoffs,
 $\varphi(\chi)=\mathrm{sgn}\chi$ and $\chi$ respectively. (Here, we
have implicitly assumed that an agent does not consider the impact
of adopting a strategy, although the excess demand is only dependent
on the adopted ones.) Let $\Omega_{a}(t)$ be the cumulative payoff
of strategy \textit{a} at time \textit{t}. Then its updating
dynamics is described by
\begin{eqnarray}
\Omega_{a}(t+1)=\Omega_{a}(t)-\xi_{a}^{\mu^{*}(t)}\varphi(\sqrt{N}A(t)).
\end{eqnarray}
Diversity of initial preferences of strategies is introduced by
adding random biases $\omega_{ia}$ to the cumulative payoffs of
strategy \textit{a} (\textit{a}$=2,...,s$) of agent \textit{i} with
respect to her first one. The biases are drawn from a Gaussian or
binomial distribution with mean 0 and variance \textit{R}. The ratio
$\rho\equiv\textit{R}/N$ is referred to as the \textit{diversity}.

 To monitor the mutual adaptive behavior of the population, we measure the variance $\sigma^{2}/N$
 of the population making decision 1, defined by
\begin{eqnarray}
\frac{\sigma^{2}}{N}\equiv\frac{N}{4}\langle[A^{\mu^{*}(t)}(t)-\langle\mathrm{A}^{\mu^{*}(t)}(t)\rangle]^{2}\rangle
\end{eqnarray}
 where the average is taken over time when the system reaches
 the steady state, and over the random distribution of strategies
 and biases.

\section{Dynamical Transitions}

\begin{figure}
\centerline{\hspace{-0.5cm}\epsfig{figure=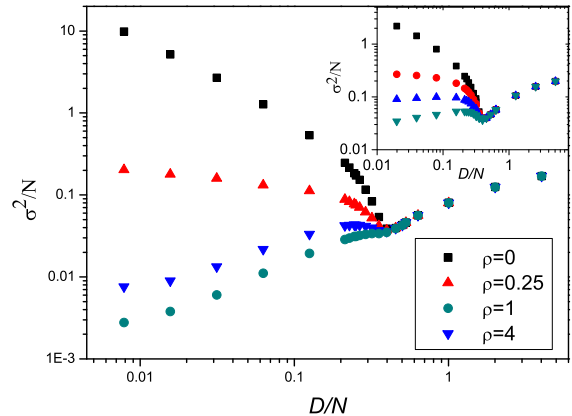,
width=1.02\linewidth}} \vspace{-0.8cm} \caption{$\sigma^2/N$ versus
$\alpha$ with linear payoffs, for $\rho$=0, 0.25, 1, 4,
respectively, $N$=251, $s$=2, 1000 samples. Inset: The corresponding
plot for step payoffs.}
\end{figure}

As shown in Fig.~1, the dependence of the variance $\sigma^{2}/N$ on
the complexity $\alpha$ for linear payoffs is very similar to that
for step payoffs \cite{Wong2004,Wong2005}. For $\alpha$ above a
universal critical value $\alpha_c(\approx 0.3)$, the variance drops
when $\alpha$ is reduced. The effects of introducing the diversity
is also similar to that for step payoffs, namely, the variance
remains unaffected when $\alpha>\alpha_{c}$, but decreases
significantly with the diversity when $\alpha<\alpha_{c}$.

However, there are differences when one goes beyond this general
trend. As shown in Fig.~2, the variance curves at different values
of $\alpha$ cross at at $\rho=\rho_c\approx0.16$, indicating the
existence of a continuous phase transition at $\rho_c$ from a phase
of vanishing variance at large $\rho$ to a phase of finite variance
at small $\rho$.

\begin{figure}
\vspace{-0.8cm} \centerline{\hspace{-0.5cm}\epsfig{figure=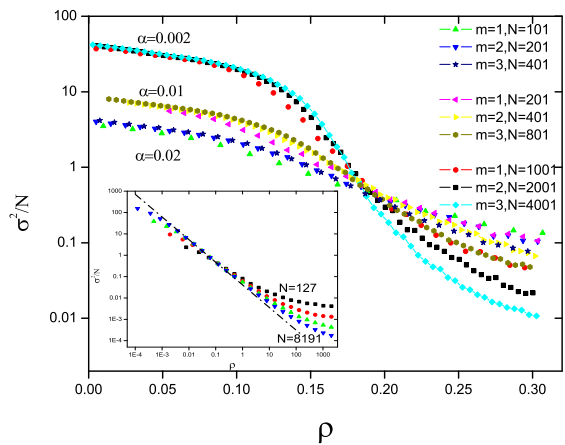,
width=1.02\linewidth}} \vspace{-0.8cm} \caption{$\sigma^2/N$ versus
$\rho$ with linear payoffs, $\alpha=0.002, 0.01, 0.02$ for different
$m$ and $N$; Inset: $\sigma^2/N$ versus $\rho$ with step payoffs
with $m=1$, $N=127, 511, 2047, 8191$ respectively. Dashed-dotted
line: scaling prediction. For both payoffs, s=2, 1000 samples. }
\end{figure}

This behavior is very different from that for step payoffs, where
the variance scales as $\rho^{-1}$ and there are no dynamical
transitions (Fig.~2 inset). The picture is confirmed by analyzing
the dynamics of the game for small $m$. The dynamics can be
conveniently described by introducing the $D$-dimensional vector
$A^{\mu}(t)\equiv\sum_{i}\xi_{a^{*}(i,t)}^{\mu}/N$. While only one
of the $D$ signals corresponds to the historical signal $\mu^{*}(t)$
of the game, the augmentation to $D$ components is necessary to
describe the attractor structure of the game dynamics. Fig.~3
illustrates the attractor structure in this phase space for the
visualizable case of $m=1$. The dynamics proceeds in the direction
which tends to reduce the magnitude of the components of
$A^{\mu}(t)$ \cite{Challet2000}. However, the components of
$A^{\mu}(t)$ overshoot, resulting in periodic attractors of period
$2D$. For $m=1$, the attractor is described by the sequence
$\mu^{*}(t)=0,1,1,0$, and takes the L-shape as shown in Fig.~3
\cite{Wong2005}. Note that the displacements in the two directions
may not have the same amplitude.

\begin{figure}
\leftline{\hspace{-0.25cm}\epsfig{figure=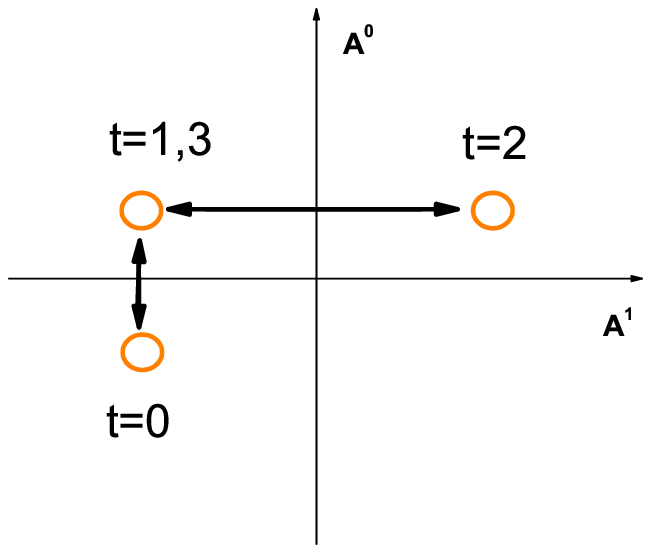,
width=0.53\linewidth}
\hspace{-0.3cm}\leftline{\epsfig{figure=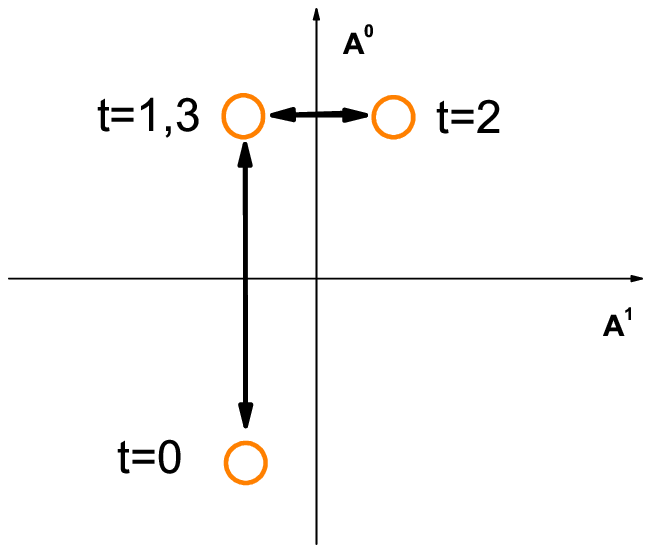,
width=0.53\linewidth}}} \vspace{-0.5cm} \caption{The attractor
dynamics when (a) $|\Delta\textit{A}^{1}|$ is larger, (b)
$|\Delta\textit{A}^{0}|$ is larger.}
\end{figure}

Following steps similar to those in \cite{Wong2005}, we find that
for $m$ not too large, and for convergence within time steps much
less than $\sqrt{R}$,
\begin{eqnarray}
A^{\mu}(t+1)=A^{\mu}(t)-\sqrt{\frac{2}{\mathrm{\pi}R}}\varphi(\sqrt{N}A^{\mu}(t))\delta_{\mu\mu^{*}(t)}.
\label{step}
\end{eqnarray}
For step payoffs, Eq.~(\ref{step}) converges to an attractor
confined in a $D$-dimensional hypercube of size
$\sqrt{2/{\mathrm{\pi}R}}$, irrespective of the value of $R$. On the
other hand, for linear payoffs, $A^{\mu}(t+1)$ becomes a linear
function of $A^{\mu}(t)$ with a slope of $1-\sqrt{2/\pi\rho}$.
Hence, for $\rho>\rho_{c}=1/{2\pi}\sim0.16$, the step sizes
$\mid$A$^{\mu}(t+1)-A^{\mu}(t)\mid$ converge to zero, whereas for
$\rho<\rho_{c}$, steps of vanishing sizes become unstable, resulting
in a continuous dynamical transition at $\rho_{c}$.

\section{The Phase of Finite Variance}

However, when $\rho<\rho_{c}$, the step sizes for each of the $D$
signals may not be equal. To see this, we monitor the variance for
each of the $D$ signals and rank them. The \textit{r}th maximum
variance is then given by
\begin{eqnarray}
S_{r}=\mathrm{large}_{\mu}\left(\frac{N}{4}[\langle(\mathrm{A}^\mathrm{\mu})^2\rangle|_{\mu=\mu^{*}(t)}-(\langle\mathrm{A}^\mathrm{\mu}\rangle|_{\mu=\mu^{*}(t)})^2],r\right)
\end{eqnarray}
where $\mathrm{large}_{\mu}(f(\mu),r)$ is the \textit{r}th largest
function $f(\mu)$ for $\mu=1,...,D$.

\begin{figure}
\centerline{\hspace{-0.5cm}\epsfig{figure=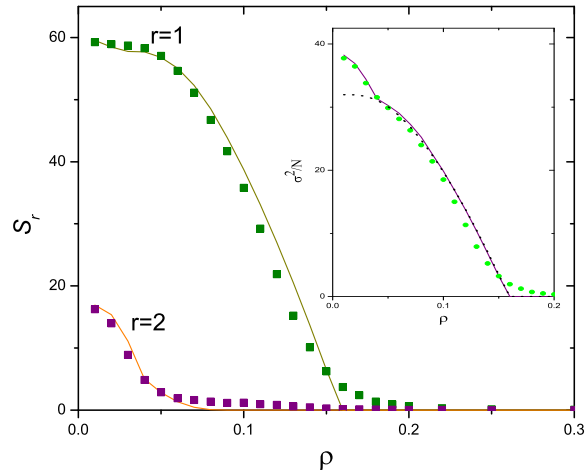,
width=1\linewidth}} \vspace{-0.8cm} \caption{$S_{r}$ versus $\rho$;
Inset: $\sigma^{2}/N$ versus $\rho$. Symbols: simulation; dotted
line: theory with one bifurcation; solid lines: theory with two
bifurcations. $N=1001$, $m=1$, $s=2$, 1000 samples.}
\end{figure}

\begin{figure}
\vspace{-0.3cm}
\leftline{\hspace{-0.5cm}\epsfig{figure=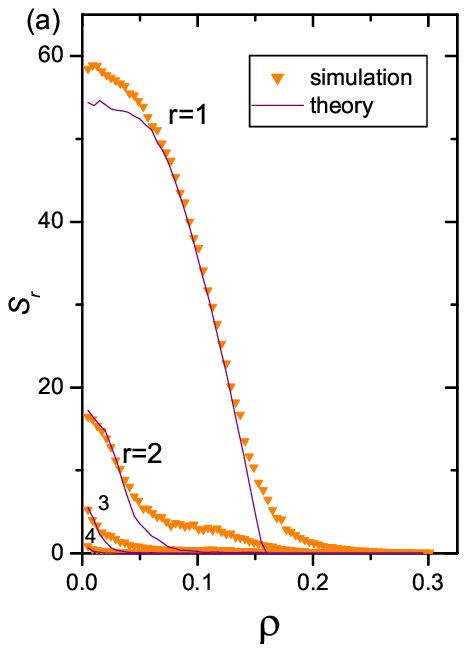,width=0.55\linewidth}
\leftline{\hspace{-0.5cm}\epsfig{figure=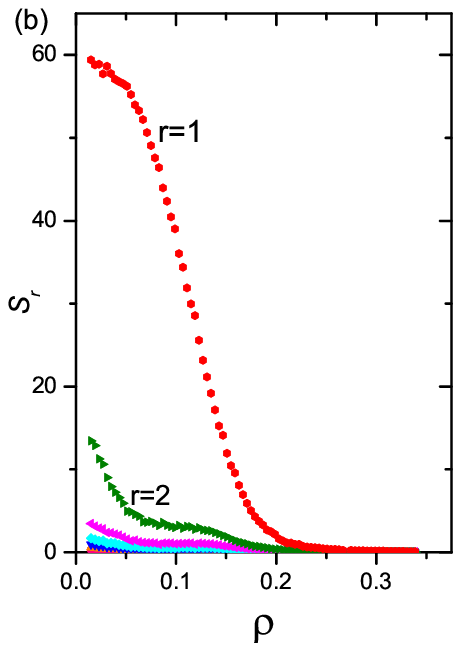,width=0.55\linewidth}}}
\vspace{-0.8cm} \caption{$S_{r}$ versus $\rho$ for (a) $m=2$, (b)
$m=3$. In both cases, $N=1001$, $s=2$, 1000 samples.}
\end{figure}

As shown in Figs.~4-5, the step sizes for each of the $D$ signals do
not bifurcate simultaneously at $\rho=\rho_{c}$, Rather, only their
first maximum bifurcates from zero when $\rho$ falls below
$\rho_{c}$, while the step sizes for the remaining $D$-1 signals
remain small. When the diversity further decreases to around 0.05,
the second maximum becomes unstable as well, and a further
bifurcation takes place. For $m\geq2$, there are further
bifurcations of the third or higher order maxima, resulting in a
\textit{cascade} of dynamical transitions when the diversity
decreases.

This cascade of transitions is confirmed by analysis. For $m=1$, we
can generalize Eq.~(\ref{step}) to convergence times of the order
$\sqrt{R}$. Assuming without loss of generality that
$\mathrm{A}^{1}$ bifurcates while $\mathrm{A}^{0}$ remains small,
 the variance of the buyer population, as derived in \cite{Ting2004}, is
\begin{eqnarray}
\frac{\sigma^{2}}{N}=\frac{N}{32}(\Delta\mathrm{A}^{1})^{2},
\quad\Delta\mathrm{A}^{1}=\mathrm{erf}\left(\frac{\Delta\mathrm{A}^{1}}{\sqrt{8\rho}}\right),
\end{eqnarray}
where $\Delta\mathrm{A}^{1}$ is the step size responding to signal
1. As Fig.~4 inset shows, the analytical and simulation results well
agree down to $\rho \sim 0.05$. However, when the diversity
decreases further, this simple analysis implies that the variance
will saturate to a constant $N/32$, whereas simulation results are
clearly higher.

This discrepancy is due to a further bifurcation of the minimum step
size. This can be analyzed by considering the effect of a
perturbation $\delta\mathrm{A}^{0}(t)$ in the direction of
$\mathrm{A}^{0}$. After a period of 4 steps, the accumulated
perturbation becomes
\begin{eqnarray}
\delta\mathrm{A}^{0}(t+4)=\left[1-\frac{1}{\sqrt{2\pi\rho}}(1+e^{-\frac{(\Delta\mathrm{A}^{1})^{2}}{8\rho}})\right]^{2}\delta\mathrm{A}^{0}(t)\label{stability}.
\end{eqnarray}
At $\rho=0.0459$, where $\Delta$A$^{1} = 0.9775$, the coefficient on
the right hand side of Eq.~(\ref{stability}) reaches the value 1,
and $\delta\mathrm{A}^{0}(t)$ diverges on further reduction of
$\rho$. Numerical iterations of the analytical equations for
$\mathrm{A}^{\mu}(t)$, averaged over samples of different initial
conditions, yield the theoretical curves in Fig.~4 and inset,
agreeing very well with simulation results. Similarly, the agreement
between analytical and simulation results are satisfactory for
$m=2$.

Since the attractors have asymmetric responses to different signals,
we also study their dependence on the initial states. Letting the
system start from a certain state (say, state 1 for $m=1$) for a
given sample, Fig.~6 shows that the initial state is more likely to
have the largest step size in the attractor for $m=1$. Simulations
show that higher values of $m$ share the same trend.

\begin{figure}[t]
\leftline{\epsfig{figure=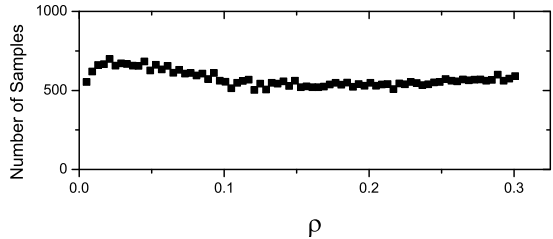,width=\linewidth}}
\vspace{-0.5cm} \caption{The number of samples with step size
responding to signal 1 being the maximum, out of 1000 samples for
$N=1001$, $m=1$, and $s=2$. The initial signal is 1. }
\end{figure}

\section{Conclusion}

We have studied the behavior of an adaptive population using a
payoff function that increases linearly with the winning margin. We
found a continuous dynamical transition when the adaptation rate of
the population is tuned by varying their diversity of preferences.
This is in contrast with the case of payoff functions independent of
the winning margin, in which no phase transitions are found.
Furthermore, we found a cascade of dynamical transitions in the
responses to different signals. This shows that an adaptive
population has the ability to self-organize to globally efficient
states and display a rich behavior, although the individual agents
make selfish decisions. Hence, despite the simplicity of the
population models, they are able to capture the essential features
of economic markets and distributed control.

\section*{Acknowledgements}
We thank S. W. Lim, and C. H. Yeung for discussions. This work is
supported by the Research Grant Council of Hong Kong
(DAG05/06.SC36).

%\end{references}

\end{document}